\documentclass{ieeetj}
\usepackage{cite}
\usepackage{amsmath,amssymb,amsfonts}
\usepackage{algorithmic}
\usepackage{graphicx}
\usepackage{textcomp}
\usepackage{xcolor}
\usepackage[nolist, nohyperlinks]{acronym}
\usepackage{orcidlink}
\usepackage{adjustbox}
\usepackage{tikz}
\usepackage{colortbl}
\usepackage{comment}

\def\BibTeX{{\rm B\kern-.05em{\sc i\kern-.025em b}\kern-.08em
    T\kern-.1667em\lower.7ex\hbox{E}\kern-.125emX}}
\AtBeginDocument{\definecolor{tmlcncolor}{cmyk}{0.93,0.59,0.15,0.02}\definecolor{NavyBlue}{RGB}{0,86,125}}

\begin{acronym}
\acro{AE}{autoencoder}
\acro{ASV}{automatic speaker verification}
\acro{BCE}{binary cross-entropy}
\acro{CM}{countermeasure}
\acro{CNN}{convolutional neural network}
\acro{COTS}{commercial-off-the-shelf}
\acro{CQCC}{constant-Q cepstral coefficient}
\acro{CRNN}{convolutional recurrent neural network}
\acro{DNN}{deep neural network}
\acro{DF}{deepfake}
\acro{EER}{equal error rate}
\acro{ELU}{exponential linear unit}
\acro{FCN}{fully convolutional network}
\acro{GMM}{Gaussian mixture model}
\acro{GRU}{gated recurrent unit}
\acro{IoT}{Internet of things}
\acro{LA}{logical access}
\acro{LSTM}{long short-term memory}
\acro{MVDR}{minimum variance distortionless response}
\acro{MWF}{multichannel Wiener filtering}
\acro{ReMASC}{Realistic Replay Attack Microphone Array Speech}
\acro{RNN}{recurrent neural network}
\acro{ROC}{receiving operating characteristic}
\acro{TTS}{text-to-speech}
\acro{PA}{physical access}
\acro{PFA}{probability of false alarm}
\acro{SOTA}{state-of-the-art}
\acro{STFT}{short-time Fourier transform}
\acro{VA}{voice assistant}
\acro{VC}{voice conversion}
\end{acronym}

\definecolor{Gray}{gray}{0.85}
\definecolor{red_cool}{rgb}{0.5, 0.0, 0.0}

\def\OJlogo{\vspace{-4pt}$<$Society logo(s) and publication title will appear here.$>$}
\def\seclogo{\vspace{10pt}$<$Society logo(s) and publication title will appear here.$>$}

\def\authorrefmark#1{\ensuremath{^{\textbf{#1}}}}

\begin{document}
\receiveddate{XX Month, XXXX}
\reviseddate{XX Month, XXXX}
\accepteddate{XX Month, XXXX}
\publisheddate{XX Month, XXXX}
\currentdate{XX Month, XXXX}
\doiinfo{XXXX.2022.1234567}

\markboth{}{Neri {et al.} Multi-channel Replay Speech Detection using an Adaptive Learnable Beamformer}

\title{Multi-channel Replay Speech Detection using an Adaptive Learnable Beamformer}

\author{Michael Neri\authorrefmark{1}, Member, IEEE, Tuomas Virtanen\authorrefmark{1}, Fellow, IEEE}
\affil{Faculty of Information Technology and Communication Sciences, Tampere University, Tampere, 33720 Finland}
%\affil{Department of Physics, Colorado State University, Fort Collins, CO 80523 USA}
%\affil{Electrical Engineering Department, University of Colorado, Boulder, CO 80309 USA}
\corresp{Corresponding author: Michael Neri (email: michael.neri@tuni.fi).}
\authornote{}

\begin{abstract}
Replay attacks belong to the class of severe threats against voice-controlled systems, exploiting the easy accessibility of speech signals by recorded and replayed speech to grant unauthorized access to sensitive data. In this work, we propose a multi-channel neural network architecture called M-ALRAD for the detection of replay attacks based on spatial audio features. This approach integrates a learnable adaptive beamformer with a convolutional recurrent neural network, allowing for joint optimization of spatial filtering and classification. Experiments have been carried out on the ReMASC dataset, which is a state-of-the-art multi-channel replay speech detection dataset encompassing four microphones with diverse array configurations and four environments. Results on the ReMASC dataset show the superiority of the approach compared to the state-of-the-art and yield substantial improvements for challenging acoustic environments. In addition, we demonstrate that our approach is able to better generalize to unseen environments with respect to prior studies.
\end{abstract}



\maketitle
\section{Introduction}
In recent years, the use of \acp{VA} has been increasing for human-machine interaction, employing voice as a biometric trait~\cite{Huang_IoTJ_2022}. With \acp{VA}, it is possible to control smart devices in our homes using \ac{IoT} networks and transmit sensitive information over the Internet. For this reason, malicious users started to attack audio-based data to grant unauthorized access to personal data and devices. 

Generally, there are several types of attacks that a malicious user can exploit toward an audio recording~\cite{Liu_TASLP_2023}. It is possible to perform a \ac{LA} attack, i.e., applying \ac{TTS} and/or \ac{VC} algorithms to modify the speech content or imitate the user. Lately, the \ac{DF} attack showed how compression and quantization operations to audio recordings can hide \ac{LA} artifacts.

\begin{figure}[ht!]
    \centering
    \includegraphics[width=1\linewidth]{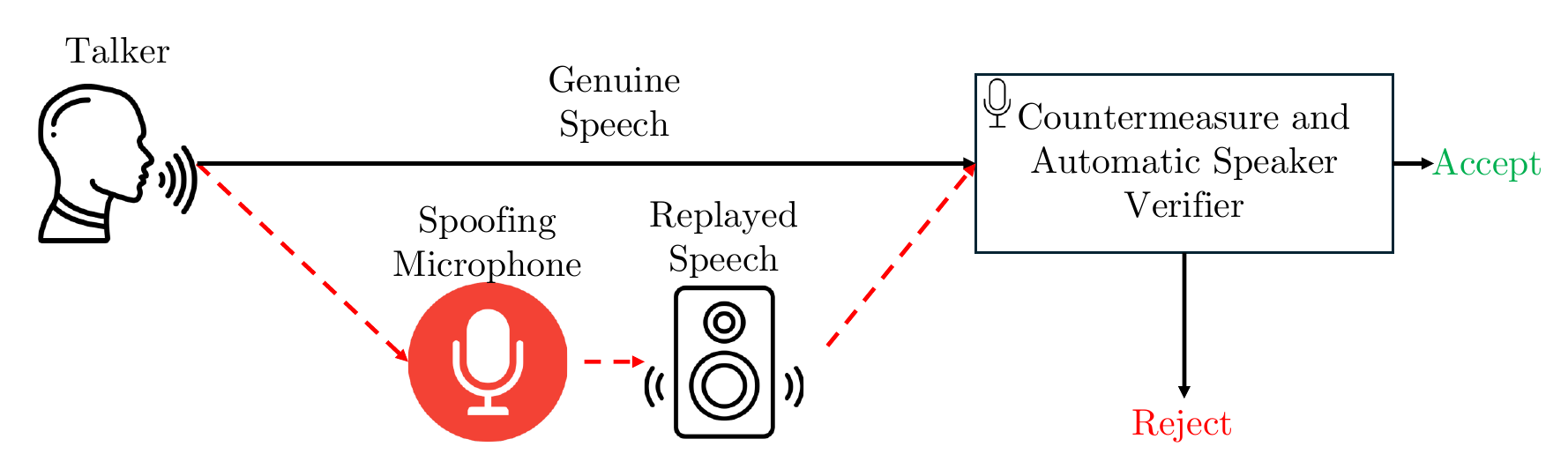}
    \caption{The problem of detecting the replay speech attack. In this scenario, a genuine speech from a talker can be either directly transmitted to a biometric system or intercepted by a spoofing microphone, recorded, and replayed through a speaker. The replayed speech, captured by the system, may be misclassified as genuine, granting unauthorized access to private data. A countermeasure and automatic speaker verification module are typically used to distinguish between genuine speech and spoofed (replayed) speech.
    }
    \label{fig:task}
\end{figure}

Differently, with \ac{PA} attacks the objective is to counterfeit the \ac{ASV} system at the microphone level. In this category, there are \textit{impersonation attacks}~\cite{Huang_IoTJ_2022} and \textit{replay attacks}~\cite{Gong_SPL_2020}. The former involves an attacker who is able to mimic the user speech of the target to grant access through the \ac{ASV} system. The latter, instead, consists of a fraudulent recording of the genuine speaker using a spoofing microphone, which is later replayed to the \ac{ASV} microphone. An example of a replay attack is shown in Figure~\ref{fig:task}.

The research community has mainly focused its attention on \ac{LA} attacks by developing \ac{CM} models that are responsible for detecting synthethic generation artifacts. In fact, several datasets have been published with the aim to tackle these attacks, e.g., the ASVSpoof series~\cite{Chettri_TASLP_2020, Liu_TASLP_2023}. However, voice is different to other biometric markers as it is always publicly exposed, easing the collection of spoofed recordings by attackers from the target speaker~\cite{Delac_ISEM_2004}. This characteristic allows malicious users to easily perform \ac{PA} even without any professional audio expertise. In fact, \textit{replay attacks} can be performed by using \ac{COTS} devices like smartphone and loudspeakers with high success rates.

In the literature, several works have tried to mitigate this type of attack by providing single-channel speech datasets, such as RedDots~\cite{Kinnunen_ICASSP_2017}, ASVSpoof2017 \ac{PA}~\cite{kinnunen2017asvspoof}, ASVSpoof2019~\cite{todisco2019asvspoof} \ac{PA}, and ASVSpoof2021 \ac{PA}~\cite{Liu_TASLP_2023}. Building upon these datasets, researchers have been deploying various methods to address single-channel replay attack detection, such as capsule networks~\cite{Luo_ICASSP_2021}, \acp{CNN}  with hand-crafted audio features combined with time-frequency representations~\cite{boyd2023voice, Xu_TASLP_2023, He_JIOT_2024}, and teacher-student architectures to reduce the computational complexity for real-time detection~\cite{Xue_LSP_2024}.

However, for speech enhancement and separation tasks, microphone arrays are usually employed in \ac{ASV} systems to exploit spatial information and improve audio quality~\cite{omologo2001speech}. Moreover, multi-channel data can be beneficial for detecting the replay detection for several reasons: (i) multi-channel recordings encompass audio spatial cues that can help the detection~\cite{Gong_Interspeech_2019, Gong_SPL_2020}, and (ii) this spatial information cannot be easily counterfeited by an attacker, differently to single-channel data where temporal and frequency cues can be manipulated to fool an \ac{ASV} system~\cite{zhang2016voicelive}. Unfortunately, single-channel replay detectors do not exploit spatial information from multi-channel recordings. In this regard, single-channel replay detectors demonstrated poor generalization capabilities across different environments and devices on multi-channel data~\cite{Gong_SPL_2020}. Moreover, there is a lack of datasets and models that are capable to collect and detect replay attacks based on multi-channel recordings. \ac{ReMASC} is the only publicly available dataset that encompasses real multi-channel audio samples, including different microphone arrays, playback devices, and environments. The baseline of this dataset was CQCC-GMM~\cite{Gong_Interspeech_2019}, which is a \ac{GMM} classifier trained on \acp{CQCC}. The authors of the database also proposed a \ac{CRNN} which acts as both a learnable and adaptive time-domain beamformer with a \ac{CRNN} replay attack classifier~\cite{Gong_SPL_2020}.

Recent works in \ac{SOTA} focused on hand-crafted single-channel features with simple classifiers (such as \acp{CNN} and \acp{GMM}) after a beamforming phase (TECC~\cite{Kotta_2020_APSIPA}, CTECC~\cite{Acharya_2021_ICASSP}, and ETECC~\cite{Patil_2022_CSL}). However, all these methods suffer from generalization capabilities, i.e., changing the acoustic properties on the test set yields random guess predictions. This problem also occurs with single-channel utterances~\cite{Xu_TASLP_2023}. In addition, performance for known environments is still not enough for production stage.

To tackle the aforementioned problems, the contributions of this work are:
\begin{itemize}
    \item The definition of a novel \textbf{M}ulti-channel \textbf{R}eplay \textbf{A}ttack \textbf{D}etector that exploits a \textbf{L}earnable time-frequency \textbf{A}daptive beamformer to classify between genuine and spoofed audio recordings, namely \textbf{M-ALRAD}. It includes a time-frequency beamformer and a \ac{CRNN} classifier. 
    \item We define a beamformer regularization during the training process to enhance the generalization capabilities of M-ALRAD to unseen environments.  
    \item Comparisons with the state-of-the-art show the effectiveness of the proposed approach in all the microphone configurations. In addition, we demonstrate that our approach is able to better generalize in environments where no a-priori knowledge is available about acoustic properties with respect to \ac{SOTA} methods.
\end{itemize}

The paper is organized as follows. Section~\ref{sec:2} introduces the proposed multi-channel replay attacks, illustrating the learnable beamformer, the \ac{DNN} that performs the classification, and the dataset under-analysis. Then, experimental results are shown in Section~\ref{sec:3}, comparing different \ac{DNN} configurations and \ac{SOTA} approaches. Finally, possible future works and conclusions are drawn in Section~\ref{sec:4}.

\section{Proposed method}\label{sec:2}
\begin{figure*}[ht!]
    \centering
    \includegraphics[width=1\linewidth]{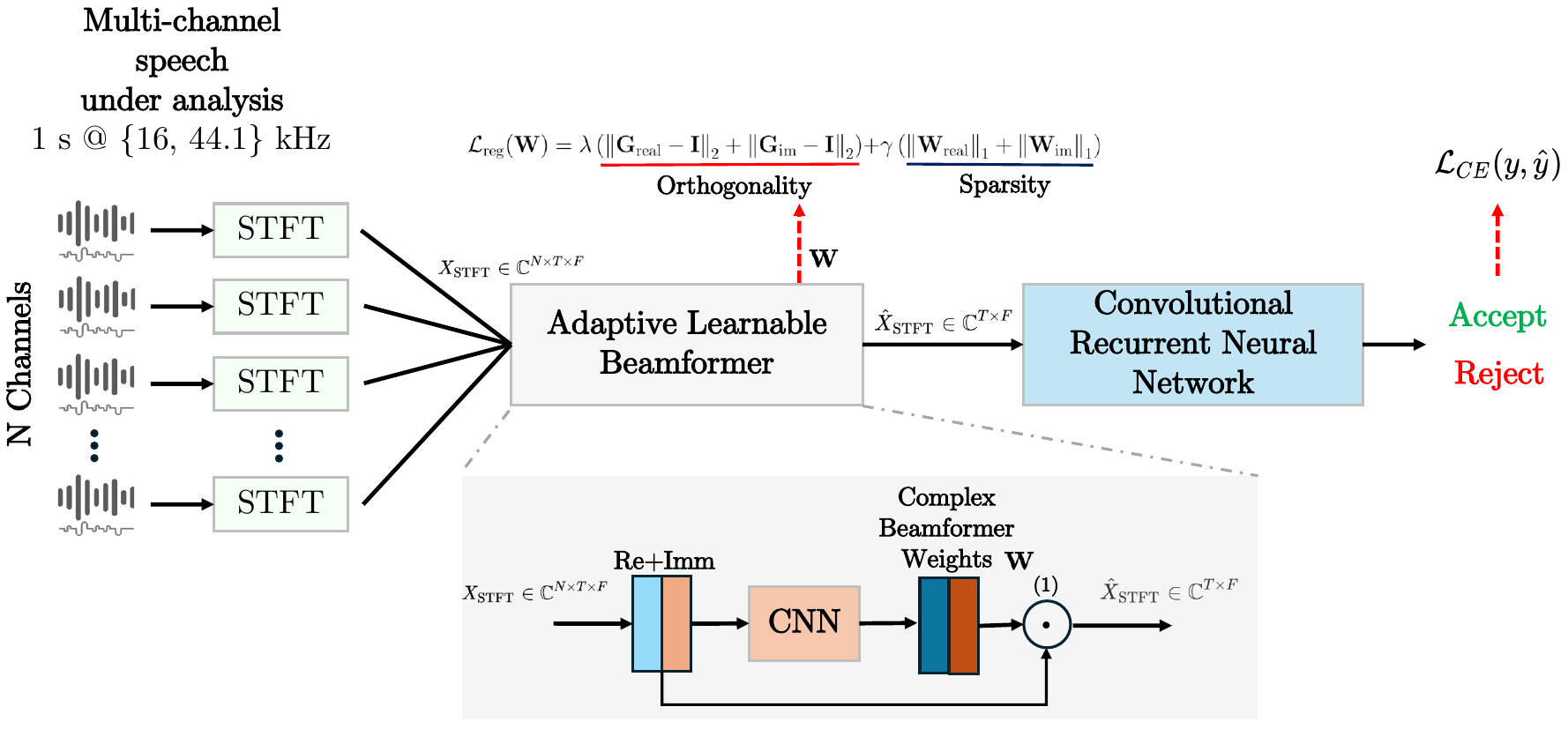}
    \caption{Overview of M-ALRAD. First, a multi-channel recording $\mathbf{x}[n]$ is channel-wise analyzed by means of \ac{STFT} to extract complex spectrograms. Then, the adaptive beamformer extracts a set of complex weights $\mathbf{W}$ that are used to obtain the final beamformed single-channel spectrogram $\hat{X}_{\mathrm{STFT}}$ . Finally, a \ac{CRNN} is employed to decide whether the recording is genuine or spoofed. 
    }
    \label{fig:overall_architecture}
\end{figure*}

In this section, the architecture of M-ALRAD is depicted. The structure of the section is as follows. First, the problem statement of this paper is shown. Then, we introduce how multi-channel recordings are processed by the adaptive learnable beamformer shown in Figure~\ref{fig:overall_architecture}. Next, we describe the \ac{DNN} that is responsible for the replay attack detection. Finally, we detail how the model is optimized during the training process.

\subsection{Problem statement}

Given a recorded speech from an array with $N$ microphones $ \mathbf{x}[n] = \{x_i[n], 0 \leq i < N\}$, the aim of this work is to identify whether the speech is genuine, i.e., produced by an human speaker, or spoofed, i.e., recorded from the target talker and then replayed again by using a loudspeaker.

\subsection{Learning-based adaptive beamforming}
Generally, beamformers are typically designed using objectives such as \ac{MVDR}~\cite{VanVeen_IEEEASSP_1988} or \ac{MWF}~\cite{brandstein2013microphone}, which are optimized to minimize noise power or mean squared error, while preserving the desired signal. While this approach suits tasks like speech enhancement and separation, where less distortion and noise improve performance, it can undermine replay attack detection by removing key cues in noisy or high-frequency components. As done in~\cite{Gong_SPL_2020}, a more effective solution is to jointly optimize the beamformer with the replay attack detector by means of the back-propagation algorithm. 

In our method, we design a learning-based adaptive beamformer. Specifically, the module is a \ac{CNN} that analyzes complex \acp{STFT} of the input multi-channel recordings to output a  single-channel beamformed complex \ac{STFT}, i.e., $f_{BM}:\mathbb{C}^{N \times T \times F} \rightarrow \mathbb{C}^{T \times F}$, where $T$ and $F$ denote the number of time and frequency bins, respectively. Next, a \ac{CNN} is defined to process all the complex spectrograms and to output the beamforming weights $\mathbf{W} \in \mathbb{C}^{N \times T \times F}$. The \ac{CNN} is a sequence of Conv2D-BatchNorm-ELU-Conv2D operations with $n_f$ $3 \times 3$ filters. Real and imaginary parts are concatenated along the channel dimension, both at the input and at the output of the learnable beamformer. Specifically, the \ac{CNN} outputs real and imaginary parts, namely $\mathbf{W}_{\text{re}} \in \mathbb{R}^{N \times T \times F}$ and $\mathbf{W}_{\text{im}} \in \mathbb{R}^{N \times T \times F}$ respectively, which are combined to compose the complex weights as $\mathbf{W} =\mathbf{W}_{\text{re}} +j\mathbf{W}_{\text{im}}$. Finally, the beamformed single-channel complex spectrogram $\hat{X}_{\mathrm{STFT}} \in \mathbb{C}^{T \times F}$ is obtained by summing all the input \acp{STFT} multiplied by the beamforming weights as

\begin{equation}
 \hat{X}_{{\mathrm{STFT}_{t, f}}} = \sum_{n} X_{{\mathrm{STFT}}_{n,t,f}} \cdot w_{n,t,f}.
\end{equation}

\subsection{Convolutional recurrent neural network classifier}
Given the single-channel complex spectrogram from the beamformer, we employ a \ac{CRNN} that has been already used for speaker distance estimation from single-channel recordings~\cite{Neri_TASLP_2024, Neri_WASPAA_2023}. Specifically, magnitude and phase of the beamformed \ac{STFT} are computed. We extract the sin\&cos features for each time-frequency point of the \ac{STFT} raw phase, i.e., $\sin (\angle\text{STFT}\{\mathbf{x}\})$ and $\cos (\angle\text{STFT}\{\mathbf{x}\})$. The features are then arranged in a $T \times F \times 3$ tensor to be processed by three convolutional layers. Each convolutional block consists of a 2D convolutional layer containing $P_i = \{P_1, P_2, P_3\} = \{32, 64, 128\}$ $1\times 3$ filters, where $i$ corresponds to the i-th layer, followed by batch normalization. Max and average pooling operation are computed in parallel along the frequency dimension in order to be summed. The applied activation function is the \ac{ELU}~\cite{Djork_2016_ICLR}. The max and average pooling rate of each layer is $MP = \{MP_1, MP_2, MP_3 \} = \{8, 8, 4\}$. Two bi-directional \ac{GRU} layers, with $\tanh(\cdot)$ as activation function, are applied to the feature maps from the convolutional layers. In this implementation, each \ac{GRU} has $128$ neurons. Finally, the hidden state of the last time step from the \ac{GRU} output is extracted, which represents the final summary of the entire sequence, and converted into the prediction $\hat{y} \in \mathbb{R}$ by means of a fully connected layer.

\subsection{Beamformer regularization and classification loss}
To increase the generalization capabilities of the proposed approach, we propose to enforce complex orthogonality and sparse weights in the learnable beamformer.

Specifically, since the matrix $W$ is 3-dimensional, we decompose the complex weights $\mathbf{W} \in \mathbb{C}^{N \times T \times F}$ into real and imaginary parts, reshaping them into $\mathbf{W}_{\text{re}} \in \mathbb{R}^{N\times F \times T}$ into $\mathbb{R}^{N \times FT}$, and similarly for $\mathbf{W}_{\text{im}}$

\begin{equation} \mathbf{W} = \mathbf{W}_{\text{re}} + j \mathbf{W}_{\text{im}}, \end{equation}
and enforce orthogonality independently on both parts:

\begin{equation} \mathcal{L}_{\text{ortho}} = \left\lVert \mathbf{W}_{\text{re}} \mathbf{W}_{\text{re}}^T - \mathbf{I} \right\rVert_2 + \left\lVert \mathbf{W}_{\text{im}} \mathbf{W}_\text{im}^T - \mathbf{I} \right\rVert_2, \end{equation}
where $\mathbf{I}$ is the identity matrix. This is done by computing their Gram matrices

\begin{equation} \mathbf{G}_{\text{re}} = \mathbf{W}_\text{re} \mathbf{W}_\text{re}^T, \qquad \mathbf{G}_{\text{im}} = \mathbf{W}_\text{im} \mathbf{W}_\text{im}^T. \end{equation}

Finally, the full regularization loss, including the sparsity constraint, is:

\begin{equation} \begin{aligned}
 \mathcal{L}_{\text{reg}}(\mathbf{W}) = \lambda \left( \left\lVert \mathbf{G}_{\text{re}} - \mathbf{I} \right\rVert_2 + \left\lVert \mathbf{G}_{\text{im}} - \mathbf{I} \right\rVert_2 \right) + \\ \gamma \left( \left\lVert \mathbf{W}_\text{re} \right\rVert_1 + \left\lVert \mathbf{W}_\text{im} \right\rVert_1 \right).
\end{aligned}\end{equation}
where $||\cdot||_{p}$ is the $p$-norm, and $\{\lambda, \gamma\}$ are used to combine losses with different magnitudes. The last term imposes sparsity of beamformer's weights, encouraging stability and prevent overly large or noisy weights in the learnable beamformer.

Finally, the overall architecture is trained to distinguish between genuine and replay recording by minimizing the cross-entropy loss $\mathcal{L}_{\mathrm{CE}}(y, \hat{y})$ between ground truth $y$ and prediction $\hat{y}$ in conjunction with the beamformer regularizer

\begin{equation}
    \mathcal{L}(y, \hat{y}, \mathbf{W}) = \mathcal{L}_{\mathrm{CE}}(y, \hat{y}) + \mathcal{L}_{reg}(\mathbf{W}).
\end{equation}

\subsection{Implementation details}
Following~\cite{Gong_SPL_2020}, we only analyze the first $1$ second of multi-channel recordings to reduce the computational complexity of the approach, as increasing the analysis segment length to $1.5$ or $2$ seconds was not found to improve the performance. Separate models are trained for each microphone array. Sampling rate for microphones $\{\mathrm{D}1, \mathrm{D}2,\mathrm{D}3\}$ is $44.1$ kHz, whereas for $\mathrm{D}4$ it is $16$ kHz. The \ac{STFT} is computed with a Hanning window of length $32$ and $46$ ms with $50 \%$ overlap for $44.1$ and $16$ kHz, respectively. The number of intermediate channels in the beamformer \ac{CNN} is $n_f = 64$. The proposed model is then trained with a batch size of $32$ using a cosine annealing scheduled learning rate of $0.001$ for $50$ epochs. We set $\lambda = \gamma = 0.00001$. These hyperparameters have been selected based on a validation set randomly extracted from the training set with a $90$-$10$ split ratio. We re-weight the cross-entropy loss for each class using the normalized reciprocal of the sample number of the class in the training set to avoid the class-imbalance problem. M-ALRAD is trained for approximately $50$ minutes on a NVIDIA GeForce RTX 4070. The average measured inference time of a single multi-channel recording is around $10$ ms. On average, M-ALRAD consists of $300$k learnable parameters. A version of the model is trained for each microphone array.   

\section{Experimental results}\label{sec:3}
In this section, we provide the results of our approach on \ac{ReMASC} dataset. In more detail, we compare the microphone-wise performance of our approach with the \ac{ReMASC} baseline. Then, we analyze the effect of varying the acoustical properties of the scenario. Finally, a comparison with prior works and an ablation study are provided.

\subsection{Evaluation}
The \ac{ReMASC} is used to assess the proposed approach~\cite{Gong_Interspeech_2019}. The dataset features a wide range of speakers, with variations in gender and vocal characteristics, providing diversity in the multi-channel audio data. It also incorporates recordings from four environments, including one outdoor scenario (Env-A), two enclosures (Env-B and Env-C) and one moving vehicle (Env-D), to capture different acoustic properties. Two spoofing microphones are employed for recording the genuine speech and then replayed using four different playback devices, including diverse quality. All recordings are synchronized across $4$ \ac{ASV} microphones $\{\mathrm{D}1, \mathrm{D}2,\mathrm{D}3,\mathrm{D}4\}$ with $2$, $4$, $6$, and $7$ channels, respectively. In total, there are $9,240$ and $45,472$ genuine and replay audio samples, respectively.

To assess the performance of our approach and directly compare with \ac{SOTA} models, \ac{EER} metric is used following the same train-test split provided by the authors of the dataset~\cite{Gong_Interspeech_2019}. To reduce the stochastic behavior of training the proposed \ac{DNN}, five independent runs are collected to compute the $95\%$ mean confidence intervals. 

\subsection{Results and comparison with baseline}\label{sec:exp_results}
Table~\ref{tab:resultsComplMicWise} shows the microphone-wise performance comparison to \ac{ReMASC} baseline NN-Multichannel~\cite{Gong_SPL_2020}.  In addition, to evaluate the effect of spatial information, we provide the results for the case where a single-channel (the first of the array) is replicated to each microphone channel and then fed to the proposed \ac{DNN}, namely ALRAD. The direct comparison between M-ALRAD and NN-Multichannel shows that our approach achieves better results across all the microphones. Notably, an average decrease of $5\%$ in terms of \ac{EER} can be highlighted, except for $\mathrm{D}1$ where the improvement is approximately $10\%$.

\begin{figure}[ht!]
    \centering
    \includegraphics[width=1\linewidth]{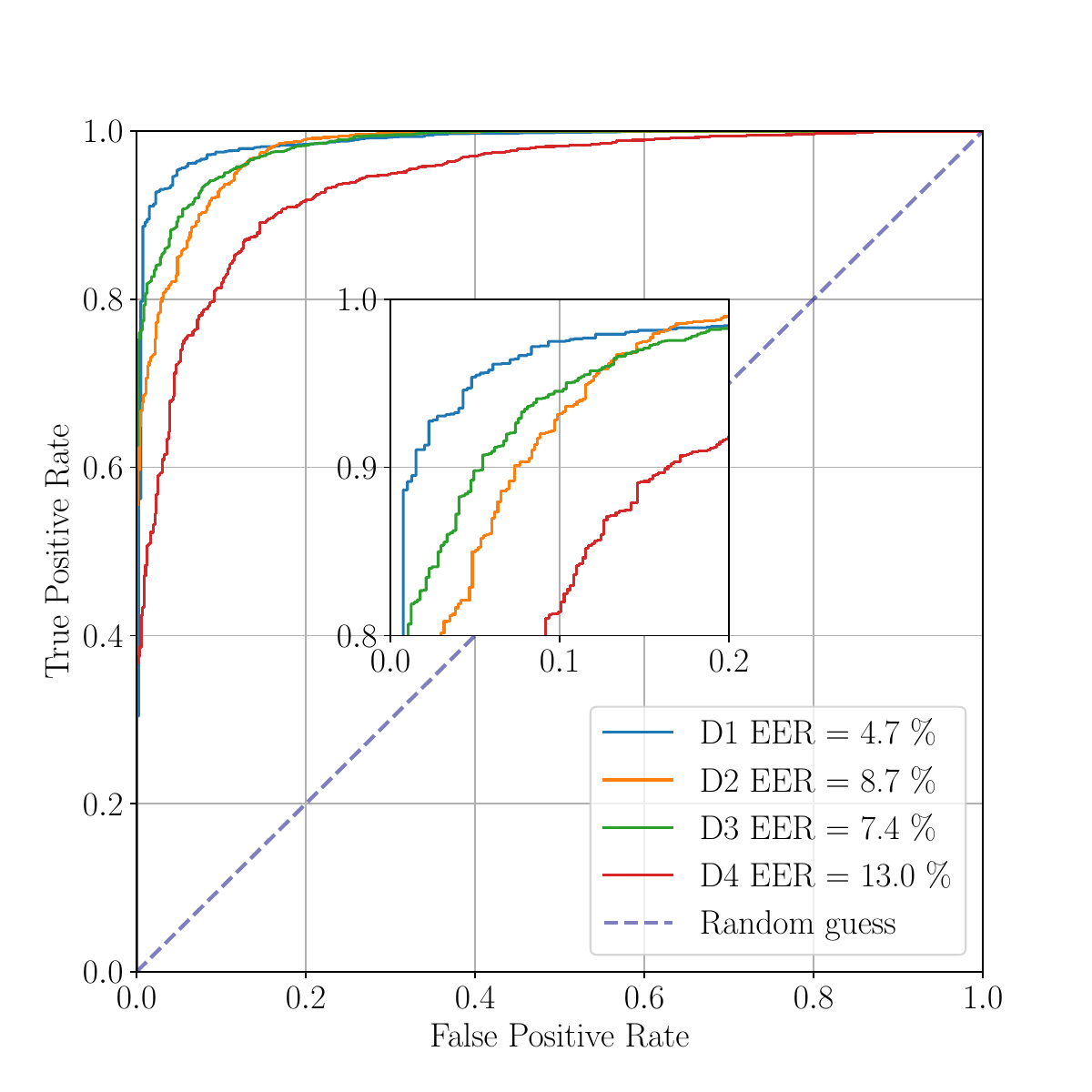}
    \caption{\ac{ROC} curves for each microphone's model.}
    \label{fig:ROCMicWise}
\end{figure}

As noted in~\cite{Gong_SPL_2020}, the use of multiple channels proves advantageous for the detection of replay attacks. However, in accordance with previous studies~\cite{Gong_Interspeech_2019, Gong_SPL_2020, Patil_2022_CSL}, it is important to note that recognizing attacks becomes more challenging with arrays that have more microphones. Specifically, while using both channels of a binaural microphone improves performance compared to a single channel, using arrays with more microphones yields worse results. To better analyze this behavior, \ac{ROC} curves of the proposed approach for each microphone are depicted in Figure~\ref{fig:ROCMicWise}.

\begin{table}[ht!]
\caption{Microphone-wise comparison of \ac{EER} ($\%$) with \ac{ReMASC}  baseline.}
\label{tab:resultsComplMicWise}
\centering
\adjustbox{max width=0.5\textwidth}{%
    \begin{tabular}{c|cccc}
    \hline \hline 
    Methods & D1 & D2 & D3 & D4  \\ 
    \hline
    NN-Single~\cite{Gong_SPL_2020} & $16.6$ & $23.7$ & $23.7$ & $27.5$ \\
    NN-Dummy Multichannel~\cite{Gong_SPL_2020} & $16.0$ & $23.2$ & $24.5$ & $25.2$ \\
    NN-Multichannel~\cite{Gong_SPL_2020} & $14.9$ & $15.4$ & $16.5$ & $19.8$ \\
    \hline
    \rowcolor{Gray} ALRAD & $5.5 \pm {\scriptstyle 2.6}$ & $11.9 \pm {\scriptstyle 2.2}$ & $19.5 \pm {\scriptstyle 2.2}$ & $21.7 \pm {\scriptstyle 2.1}$ \\
    \rowcolor{Gray} M-ALRAD & $\mathbf{5.2} \pm {\scriptstyle 1.2}$ & $\mathbf{10.0} \pm {\scriptstyle 0.9}$ & $\mathbf{10.4} \pm {\scriptstyle 2.1}$ & $\mathbf{14.2} \pm {\scriptstyle 0.8}$ \\
    \hline \hline
    \end{tabular}
    }
\end{table}

\subsection{Generalization across unseen environments}
To assess the generalization capabilities of our approach, we carried out dependent training procedures as in~\cite{Gong_Interspeech_2019}. Specifically, we assess the performance of the model when either samples of the target environment are present (\textit{environment-dependent}) or not (\textit{environment-independent}) in the training set. For example, results for \textit{environment-dependent} Env-A are obtained by training M-ALRAD for each microphone with all the environments, testing on Env-A, and averaging the \acp{EER}. Regarding \textit{environment-independent} on the same environment, Env-A is not present in the training set, enabling the evaluation of M-ALRAD to unseen scenes.

Table~\ref{tab:resultsEnv} shows a comparison between \ac{SOTA} approaches and our replay attack detector. Generally, in the case when the model has some a-priori knowledge of the environment, our approach outperforms the \ac{SOTA} approaches in the outdoor scenario and in enclosures with diffuse noise and different speaker-microphone configurations. Instead, the performance in the moving vehicle environment is slightly lower with respect to the \ac{SOTA}. 

Prior works were not able to generalize when no samples of the target environment are present in the training set. Instead, our approach is able to better recognize the replay attack in all the environment with large \ac{EER} improvements.

\begin{table}[ht!]
\caption{\acp{EER} ($\%$) for environment-dependent vs. environment-independent scenarios on ReMASC dataset.}
\label{tab:resultsEnv}
\centering
\adjustbox{max width=0.5\textwidth}{%
    \begin{tabular}{c|cccc}
    \hline \hline 
    Methods & Env-A & Env-B$^{*}$ & Env-C & Env-D  \\ 
    \hline
    \multicolumn{5}{c}{Environment-dependent} \\
    CQCC-GMM~\cite{Gong_Interspeech_2019} & $13.5$  & $17.4$  & $21.3$  & $22.1$  \\
    ETECC~\cite{Patil_2022_CSL} & $15.1$  & $33.8$  & $14.1$  & $10.4$ \\
    CTECC~\cite{Acharya_2021_ICASSP} & $13.0$  & $26.8$  & $9.9$  & $10.1$ \\
    TECC~\cite{Kotta_2020_APSIPA} & $13.4$ & $27.9$ & $10.3$ & $\mathbf{9.1}$ \\
    \rowcolor{Gray} M-ALRAD & $\mathbf{8.1} \pm {\scriptstyle 2.3}$ &  $\mathbf{5.8} \pm {\scriptstyle 2.4}$  &  $\mathbf{7.5} \pm {\scriptstyle 2.6}$ & $14.0 \pm {\scriptstyle 2.0}$  \\
    \hline
    \multicolumn{5}{c}{Environment-independent} \\
    CQCC-GMM~\cite{Gong_Interspeech_2019} & $19.9$  & $39.9$  & $34.6$  & $48.9$  \\
    ETECC~\cite{Patil_2022_CSL} & $29.0$  & $32.5$  & $30.0$  & $49.9$ \\
    CTECC~\cite{Acharya_2021_ICASSP} & $28.5$  & $34.7$  & $32.5$  & $50.0$ \\
    TECC~\cite{Kotta_2020_APSIPA} & $26.8$ & $35.4$ & $31.8$ & $50.0$ \\
    \rowcolor{Gray} M-ALRAD & $\mathbf{13.8} \pm {\scriptstyle 2.8}$ & $\mathbf{20.3} \pm {\scriptstyle 3.1}$  & $\mathbf{15.1} \pm {\scriptstyle 5.5}$ & $\mathbf{24.2} \pm {\scriptstyle 2.4}$  \\
    \hline \hline
    \multicolumn{5}{l}{\scriptsize Env-B$^{*}$ does not encompass D1 genuine utterances due to hardware fault during data collection.} \\
    \end{tabular}
    }
\end{table}

\subsection{Analysis on fixed and learnable beamformer}\label{sec:bm}
To demonstrate the effectiveness of the adaptive learnable beamformer, Table~\ref{tab:resultsCompl} shows the performance of different configurations of the beamformer, following the same experimental setup as in Section~\ref{sec:exp_results}. We analyzed the scenarios where (i) a learnable beamformer is present but not adaptive, i.e., single-channel weights $ \mathbf{W} \in \mathbb{C}^{F}$ are learnable during training and fixed during inference, (ii) same as (i) but with multi-channel weights $\mathbf{W} \in \mathbb{C}^{N \times F}$, (iii) M-ALRAD without any regularization, (iv) M-ALRAD without orthogonality regularization, and (v) M-ALRAD without sparsity constraint.

It is important to note that the fixed beamformer, i.e., with predetermined weights $\mathbf{W}$ for all audio recordings, performs similarly to the variable version when the number of channels is low ($2$ and $4$ channel microphones). However, this configuration performs worse with the arrays $\mathrm{D}3$ and $\mathrm{D}4$, with a decrease in performance of $6\%$ and $1.3\%$. This could be due to the number of microphones in these arrays, where fixed complex beamforming weights cannot adapt to more general scenarios.

In fact, using the adaptive beamformer in which the weights $\mathbf{W}$ depend on the input audio, it is possible to better generalize the task, providing better performance in scenarios where \ac{SOTA} models struggled to identify replay attacks due to varying spatial characteristics. 

Similarly, beamformer regularizations are more effective with larger microphone arrays, as it can be inspected from Table~\ref{tab:resultsCompl}. Moreover, the combination of orthogonality and sparsity constraints yielded the best performance, although small \ac{EER} improvements have been achieved by using only one regularizer.

\begin{table}[ht!]
\caption{Ablation study with respect to EER (\%).}
\label{tab:resultsCompl}
\centering
\adjustbox{max width=0.5\textwidth}{%
    \begin{tabular}{c|cccc}
    \hline \hline 
    Methods & D1 & D2 & D3 & D4  \\ 
    \hline
    (i) Learnable fixed single-channel bm & $7.3 \pm {\scriptstyle 2.6}$ & $12.3 \pm {\scriptstyle 2.5}$ & $14.6 \pm {\scriptstyle 1.1}$ & $19.9 \pm {\scriptstyle 3.1}$ \\
    (ii) Learnable fixed multi-channel bm & $5.8 \pm {\scriptstyle 1.3}$ & $11.4 \pm {\scriptstyle 4.1}$ & $16.5 \pm {\scriptstyle 1.5}$ & $15.5 \pm {\scriptstyle 2.1}$ \\
    (iii) Without any regularization & $7.3 \pm {\scriptstyle 1.8}$  & $12.2 \pm {\scriptstyle 1.7}$ & $11.2 \pm {\scriptstyle 4.5}$ & $23.0 \pm {\scriptstyle 5.2}$ \\
    (iv) Without orthogonality & $6.2 \pm {\scriptstyle 2.4}$  & $9.4 \pm {\scriptstyle 1.6}$ & $10.6 \pm {\scriptstyle 5.8}$ & $17.1 \pm {\scriptstyle 2.9}$ \\
    (v) Without sparsity & $6.6 \pm {\scriptstyle 1.0}$  & $8.7 \pm {\scriptstyle 2.2}$ & $11.2 \pm {\scriptstyle 4.5}$ & $17.8 \pm {\scriptstyle 4.5}$ \\
    \hline
    \rowcolor{Gray} M-ALRAD & $\mathbf{5.2} \pm {\scriptstyle 1.2}$ & $\mathbf{10.0} \pm {\scriptstyle 0.9}$ & $\mathbf{10.4} \pm {\scriptstyle 2.1}$ & $\mathbf{14.2} \pm {\scriptstyle 0.8}$ \\
    \hline \hline
    \end{tabular}
    }
\end{table}

\subsection{Comparison with state-of-the-art}
Since several previous works did not provide microphone-wise performance, we report the mean \ac{EER}, namely m\ac{EER}, across the entire test set in Table~\ref{tab:resultsComplmeanEER}. From the global results, the superiority of our approach is clear. Although we were able to decrease the m\ac{EER} from $14.7\%$ to $10.0\%$, this error rate still indicates significant room for improvement with respect to other biometric traits.

\begin{table}[ht!]
\caption{Comparison of m\ac{EER} (\%) with state-of-the-art approaches on evaluation subset of ReMASC~\cite{Gong_Interspeech_2019}.}
\label{tab:resultsComplmeanEER}
\centering
\adjustbox{max width=0.4\textwidth}{%
    \begin{tabular}{c|c}
    \hline \hline 
    Methods & m\ac{EER} ($\%$) \\ 
    \hline
    NN-Single~\cite{Gong_SPL_2020} & $22.9$ \\
    NN-Dummy Multichannel~\cite{Gong_SPL_2020} & $22.2$ \\
    Lexa~\cite{Rodriguez_2023_AIIoT} & $18.9$ \\
    TECC~\cite{Chodingala_2022_SPCOM} & $17.9$ \\
    NN-Multichannel~\cite{Gong_SPL_2020} & $16.7$ \\
    ETECC~\cite{Patil_2022_CSL}& $15.4$ \\
    LCNN+GMM~\cite{Acharya_2021_ICASSP} & $14.0$ \\
    TECC~\cite{Kotta_2020_APSIPA} & $14.7$ \\
    \hline
    \rowcolor{Gray} M-ALRAD & $\mathbf{10.0}$  \\
    \hline \hline
    \end{tabular}
    }
\end{table}

It is important to highlight that other versions of M-ALRAD, such as those investigated in Section~\ref{sec:bm}, gave better performance from the \ac{SOTA}, except for the case where a single-channel is replicated, i.e., without exploiting spatial information. This suggests that the combination of a learnable beamformer and a \ac{CRNN} as a classifier is enabling \ac{SOTA} performance.

% \subsection{Visualization of learned beamformer patterns}

\section{Conclusions}\label{sec:4}
In this work, we proposed a novel replay attack detection, namely M-ALRAD, based on a end-to-end \ac{DNN}, which includes a learnable and adaptive beamformer and a \ac{CRNN} that exploits spatial cues from the beamformed signal. Specifically, it performs both time-frequency beamforming and detection from multi-channel recordings. Experimental results on the \ac{ReMASC} dataset showed the superiority of our approach with respect to the \ac{SOTA}, reducing the m\ac{EER} by $4.7\%$. In addition, we designed a single-channel version, namely ALRAD, to prove the efficacy to exploit spatial cues in detecting replay speech attack. Finally, we demonstrated that the use of an adaptive time-frequency beamformer with multi-channel weights is helpful to tackle the replay speech attack both in seen and unseen scenarios, providing better generalization capabilities with respect to the \ac{SOTA}. The analysis of microphone array mismatches, including different number of geometries and microphones, for replay speech detection using multi-channel recordings is set as future work.

\bibliographystyle{IEEEtran}
\bibliography{biblio}

\begin{thebibliography}{10}
\providecommand{\url}[1]{#1}
\csname url@samestyle\endcsname
\providecommand{\newblock}{\relax}
\providecommand{\bibinfo}[2]{#2}
\providecommand{\BIBentrySTDinterwordspacing}{\spaceskip=0pt\relax}
\providecommand{\BIBentryALTinterwordstretchfactor}{4}
\providecommand{\BIBentryALTinterwordspacing}{\spaceskip=\fontdimen2\font plus
\BIBentryALTinterwordstretchfactor\fontdimen3\font minus \fontdimen4\font\relax}
\providecommand{\BIBforeignlanguage}[2]{{%
\expandafter\ifx\csname l@#1\endcsname\relax
\typeout{** WARNING: IEEEtran.bst: No hyphenation pattern has been}%
\typeout{** loaded for the language `#1'. Using the pattern for}%
\typeout{** the default language instead.}%
\else
\language=\csname l@#1\endcsname
\fi
#2}}
\providecommand{\BIBdecl}{\relax}
\BIBdecl

\bibitem{Huang_IoTJ_2022}
W.~Huang, W.~Tang, H.~Jiang, J.~Luo, and Y.~Zhang, ``Stop deceiving! an effective defense scheme against voice impersonation attacks on smart devices,'' \emph{IEEE Internet of Things Journal}, vol.~9, no.~7, pp. 5304--5314, 2022.

\bibitem{Liu_TASLP_2023}
X.~Liu, X.~Wang, M.~Sahidullah, J.~Patino, H.~Delgado, T.~Kinnunen, M.~Todisco, J.~Yamagishi, N.~Evans, A.~Nautsch, and K.~A. Lee, ``{ASVspoof 2021: Towards Spoofed and Deepfake Speech Detection in the Wild},'' \emph{IEEE/ACM Transactions on Audio, Speech, and Language Processing}, vol.~31, pp. 2507--2522, 2023.

\bibitem{Gong_SPL_2020}
Y.~Gong, J.~Yang, and C.~Poellabauer, ``{Detecting replay attacks using multi-channel audio: A neural network-based method},'' \emph{IEEE Signal Processing Letters}, vol.~27, pp. 920--924, 2020.

\bibitem{Chettri_TASLP_2020}
B.~Chettri, E.~Benetos, and B.~L.~T. Sturm, ``{Dataset Artefacts in Anti-Spoofing Systems: A Case Study on the ASVspoof 2017 Benchmark},'' \emph{IEEE/ACM Transactions on Audio, Speech, and Language Processing}, vol.~28, pp. 3018--3028, 2020.

\bibitem{Delac_ISEM_2004}
K.~Delac and M.~Grgic, ``A survey of biometric recognition methods,'' in \emph{Proceedings. Elmar-2004. 46th International Symposium on Electronics in Marine}, 2004.

\bibitem{Kinnunen_ICASSP_2017}
T.~Kinnunen, M.~Sahidullah, M.~Falcone, L.~Costantini, R.~G. Hautamäki, D.~Thomsen, A.~Sarkar, Z.~Tan, H.~Delgado, M.~Todisco, N.~Evans, V.~Hautamäki, and K.~A. Lee, ``{RedDots replayed: A new replay spoofing attack corpus for text-dependent speaker verification research},'' in \emph{IEEE International Conference on Acoustics, Speech and Signal Processing (ICASSP)}, 2017.

\bibitem{kinnunen2017asvspoof}
T.~Kinnunen, M.~Sahidullah, H.~Delgado, M.~Todisco, N.~Evans, J.~Yamagishi, and K.~A. Lee, ``{The ASVspoof 2017 challenge: Assessing the limits of replay spoofing attack detection},'' in \emph{Interspeech}, 2017.

\bibitem{todisco2019asvspoof}
M.~Todisco, X.~Wang, V.~Vestman, M.~Sahidullah, H.~Delgado, A.~Nautsch, J.~Yamagishi, N.~Evans, T.~Kinnunen, and K.~A. Lee, ``{ASVspoof 2019: Future horizons in spoofed and fake audio detection},'' in \emph{Interspeech}, 2019.

\bibitem{Luo_ICASSP_2021}
A.~Luo, E.~Li, Y.~Liu, X.~Kang, and Z.~J. Wang, ``{A Capsule Network Based Approach for Detection of Audio Spoofing Attacks},'' in \emph{IEEE International Conference on Acoustics, Speech and Signal Processing (ICASSP)}, 2021.

\bibitem{boyd2023voice}
J.~Boyd, M.~Fahim, and O.~Olukoya, ``{Voice spoofing detection for multiclass attack classification using deep learning},'' \emph{Machine Learning With Applications}, vol.~14, p. 100503, 2023.

\bibitem{Xu_TASLP_2023}
L.~Xu, J.~Yang, C.~H. You, X.~Qian, and D.~Huang, ``{Device Features Based on Linear Transformation With Parallel Training Data for Replay Speech Detection},'' \emph{IEEE/ACM Transactions on Audio, Speech, and Language Processing}, vol.~31, pp. 1574--1586, 2023.

\bibitem{He_JIOT_2024}
R.~He, Y.~Cheng, Z.~Zheng, X.~Ji, and W.~Xu, ``{Fast and Lightweight Voice Replay Attack Detection via Time-Frequency Spectrum Difference},'' \emph{IEEE Internet of Things Journal}, vol.~11, no.~18, pp. 29\,798--29\,810, 2024.

\bibitem{Xue_LSP_2024}
J.~Xue, C.~Fan, J.~Yi, J.~Zhou, and Z.~Lv, ``{Dynamic Ensemble Teacher-Student Distillation Framework for Light-Weight Fake Audio Detection},'' \emph{IEEE Signal Processing Letters}, vol.~31, pp. 2305--2309, 2024.

\bibitem{omologo2001speech}
M.~Omologo, M.~Matassoni, and P.~Svaizer, ``{Speech recognition with microphone arrays},'' in \emph{Microphone arrays: signal processing techniques and applications}.\hskip 1em plus 0.5em minus 0.4em\relax Springer, 2001, pp. 331--353.

\bibitem{Gong_Interspeech_2019}
Y.~Gong, J.~Yang, J.~Huber, M.~MacKnight, and C.~Poellabauer, ``{ReMASC: Realistic Replay Attack Corpus for Voice Controlled Systems},'' \emph{Interspeech}, 2019.

\bibitem{zhang2016voicelive}
L.~Zhang, S.~Tan, J.~Yang, and Y.~Chen, ``{Voicelive: A phoneme localization based liveness detection for voice authentication on smartphones},'' in \emph{Proceedings of the 2016 ACM SIGSAC Conference on Computer and Communications Security}, 2016, pp. 1080--1091.

\bibitem{Kotta_2020_APSIPA}
H.~Kotta, A.~T. Patil, R.~Acharya, and H.~A. Patil, ``{Subband Channel Selection using TEO for Replay Spoof Detection in Voice Assistants},'' in \emph{Asia-Pacific Signal and Information Processing Association Annual Summit and Conference (APSIPA ASC)}, 2020.

\bibitem{Acharya_2021_ICASSP}
R.~Acharya, H.~Kotta, A.~T. Patil, and H.~A. Patil, ``{Cross-Teager Energy Cepstral Coefficients for Replay Spoof Detection on Voice Assistants},'' in \emph{IEEE International Conference on Acoustics, Speech and Signal Processing (ICASSP)}, 2021, pp. 6364--6368.

\bibitem{Patil_2022_CSL}
A.~T. Patil, R.~Acharya, H.~A. Patil, and R.~C. Guido, ``{Improving the potential of enhanced teager energy cepstral coefficients (ETECC) for replay attack detection},'' \emph{Computer Speech \& Language}, vol.~72, p. 101281, 2022.

\bibitem{VanVeen_IEEEASSP_1988}
B.~Van~Veen and K.~Buckley, ``{Beamforming: a versatile approach to spatial filtering},'' \emph{IEEE ASSP Magazine}, vol.~5, no.~2, pp. 4--24, 1988.

\bibitem{brandstein2013microphone}
M.~Brandstein and D.~Ward, \emph{{Microphone arrays: signal processing techniques and applications}}.\hskip 1em plus 0.5em minus 0.4em\relax Springer Science \& Business Media, 2013.

\bibitem{Neri_TASLP_2024}
M.~Neri, A.~Politis, D.~A. Krause, M.~Carli, and T.~Virtanen, ``{Speaker Distance Estimation in Enclosures From Single-Channel Audio},'' \emph{IEEE/ACM Transactions on Audio, Speech, and Language Processing}, vol.~32, pp. 2242--2254, 2024.

\bibitem{Neri_WASPAA_2023}
M.~Neri, A.~Politis, D.~A. Krause, M.~Carli, and T.~Virtanen, ``{Single-Channel Speaker Distance Estimation in Reverberant Environments},'' in \emph{IEEE Workshop on Applications of Signal Processing to Audio and Acoustics (WASPAA)}, 2023.

\bibitem{Djork_2016_ICLR}
C.~Djork-Arné, T.~Unterthiner, and S.~Hochreiter, ``{Fast and Accurate Deep Network Learning by Exponential Linear Units (ELUs)},'' in \emph{International Conference on Learning Representations (ICLR)}, 2016.

\bibitem{Rodriguez_2023_AIIoT}
R.~Rodriguez and Y.~Li, ``{Lexa: A Liveness Detection Enabled Voice Assistant},'' in \emph{IEEE World AI IoT Congress (AIIoT)}, 2023.

\bibitem{Chodingala_2022_SPCOM}
P.~K. Chodingala, S.~S. Chaturvedi, A.~T. Patil, and H.~A. Patil, ``Robustness of das beamformer over mvdr for replay attack detection on voice assistants,'' in \emph{IEEE International Conference on Signal Processing and Communications (SPCOM)}, 2022.

\end{thebibliography}

\end{document}